\documentclass[doublecol]{epl2}

\usepackage{color,epsfig,rotating,graphicx,subfigure}
\usepackage{amsmath,amssymb,amscd}

\usepackage[latin1]{inputenc}
\usepackage[english]{babel}

\usepackage{dsfont}
\usepackage{textcomp}

\usepackage{ulem,stmaryrd}

\makeatletter
\newsavebox{\@brx}
\newcommand{\llangle}[1][]{\savebox{\@brx}{\(\m@th{#1\langle}\)}%
  \mathopen{\copy\@brx\mkern2mu\kern-0.9\wd\@brx\usebox{\@brx}}}
\newcommand{\rrangle}[1][]{\savebox{\@brx}{\(\m@th{#1\rangle}\)}%
  \mathclose{\copy\@brx\mkern2mu\kern-0.9\wd\@brx\usebox{\@brx}}}
\makeatother

\renewcommand{\Re}{{\rm Re}}
\renewcommand{\Im}{{\rm Im}}

\newcommand{\ri}{{\rm i}}

\newcommand{\rd}{{\rm d}}

\newcommand{\kb}{k_{\rm B}}

\newcommand{\Tr}{{\rm Tr}}

\newcommand{\domega}{\mathrm{d} \omega}

\title{General trace formula for heat flux fluctuations}

\author{F. Herz \and C. Kathmann \and S.-A. Biehs}

\institute{                    
  \inst{} Institut f\"{u}r Physik, Carl von Ossietzky Universit\"{a}t, D-26111 Oldenburg, Germany\\
}
\abstract{
Within the framework of macroscopic quantum electrodynamics and scattering theory, we derive the general expressions for
the variance of radiative heat transfer between two arbitrarily shaped objects placed in an arbitrary environment in 
terms of their T-operators. The such derived expression is valid in the far- and near-field regime of thermal radiation 
as well as for reciprocal and non-reciprocal objects of any size within a reciprocal or non-reciprocal environment 
as long as the distances between the objects and their sizes are covered by the macroscopic approach. As special cases
 we discuss the variance of the radiative heat flux between two nanoparticles and between a nanoparticle and a substrate in near- and far-field regime.
}

\begin{document}
\maketitle
%\tableofcontents
%\newpage

%%%%%%%%%%%%%%%%%%%%%%%%%%%%%%%%%%%%%%%%%%%%%%%%%%%%%%%%%%%%%%%%%%%%%%%%%%%%%%%%%%%%
%
% Introduction
%
%%%%%%%%%%%%%%%%%%%%%%%%%%%%%%%%%%%%%%%%%%%%%%%%%%%%%%%%%%%%%%%%%%%%%%%%%%%%%%%%%%%%

\section{Introduction}

It has been shown by several theoretical and experimental works that the Stefan-Boltzmann law does not 
provide an upper limit of thermal radiation (TR) between two media for distances smaller than the thermal wavelength anymore. 
This is due to the fact that the Stefan-Boltzmann law only takes propagating modes into account, whereas for distances 
smaller than the thermal wavelength evanescent waves will contribute to the energy flux as well~\cite{PvH,Joulain2,Volokitin}. These evanescent modes, and in particular surface modes, increase the heat flux (HF) and can result in HFs, which can 
be orders of magnitude larger than the blackbody value~\cite{PvH,Joulain2,Volokitin} as whitnessed by several recent 
experimental setups for a plane-plane geometry~\cite{Ottens,Hu,Kralik,Lim,Watjen,Bernadi,Song,Fiorino} or a sphere-plane geometry~\cite{Shen,Rousseau,Zwol2,Zwol3,Shi,Gotsmann,Kim,Cui} for polar materials, metals, phase-change materials, and hyperbolic materials. Consequently, new theoretical limits for the near-field HF are needed and have been derived~\cite{PBAlimit,BiehsEtAl2010,BiehsEtAl2012,Miller2015,Venkataram}. Recently, it could also be shown theoretically and experimentally that, when making the thickness of two opposite membranes much smaller than the thermal wavelength, then even in the far-field regime the  HF is not limited by the Stefan-Boltzmann law~\cite{Hurtado,Thompson}. In principle, this effect is known for nanoparticles (NPs) for a long time and can be traced back to the fact that the absorption cross section of a NP can be larger than its surface~\cite{Biehs}. 

These new theoretical insights into the properties of TR at the nanoscale can be exploited for  
near-field thermal imaging methods~\cite{Babuty2013,Kittel2008,Worbes2013,Jones2012,Weng2018,Komiyama2019} and energy 
harvesting by near-field thermophotovoltaic devices~\cite{Narayanaswamy,ReddyNTPV}, for instance. Furthermore, new possibilities for passive and active thermal management at the nanoscale using the specific properties of phase-change or 
non-reciprocal magneto-optical materials have been presented, not only leading to interesting concepts like near-field HF diodes~\cite{PBASAB2013,LipingWang,ItoEtAl,FiorinoEtAl2018,OttDiode}, transistors~\cite{PBASAB2014}, memories~\cite{Kubytskyi2014,DyakovEtAl2015}, logic gates~\cite{PBASAB2016,Kathmann}, but also to new fundamental effects 
for TR like a persistent HF~\cite{Zhu2016,Silveirinha, OttEtAl2018}, persistent angular momentum and spin~\cite{Silveirinha, OttEtAl2018, Zubin2019}, giant magnetic resistance~\cite{Latella2017,Cuevas}, and a 
Hall effect for TR~\cite{Ben-Abdallah2016,OttHall}. However, it is interesting to note that all these findings are related to the first-order coherence properties of TR, only. When considering, for example, the HF, then it is in most cases characterized by the mean values of the Poynting vector, i.e.\ by the correlation function $\llangle \mathbf{E} \times \mathbf{H} \rrangle$ for the fluctuational thermal electric and magnetic 
fields $\mathbf{E}$ and $\mathbf{H}$ where $\llangle \circ \rrangle$ symbolizes the ensemble average. In contrast, 
higher-order coherence properties like the variance of the HF, for example, which is connected to correlation functions of the form $\llangle (\mathbf{E} \times \mathbf{H})^2 \rrangle$, have only scarcely been studied in the near-field regime. There are only a few works discussing, for example, the fluctuations of radiative heat transport in the near-field regime~\cite{BiehsFluct,Tang}, or deriving Green-Kubo relations for TR, and connecting by this the equilibrium fluctuations of second-order to the linear transport coefficients in reciprocal and non-reciprocal systems~\cite{GreenKubo,GreenKubo2}. Note that the second-order correlation functions are also needed to determine vacuum friction~\cite{ZuritaEtAl2004} and Casimir force fluctuations~\cite{WuEtAl2002}. 

It could, for example, be shown within the framework of fluctuational electrodynamics that the standard deviation of the Poynting vector describing the TR emitted by a black body is on the same order as the mean value itself~\cite{MandelWolf} so that $\sigma_{\rm BB} \approx \langle S \rangle$. This is in agreement with the standard blackbody theory yielding a standard deviation of $\sigma_{I,\rm BB} = \langle I \rangle/\sqrt{2}$ for the mean intensity $\langle I \rangle$ of an unpolarized beam of TR, i.e.\ we have $\sigma_{I,\rm BB} \approx \langle I \rangle$.  Similar results have been obtained for the Casimir force. The stress tensor is due to the vacuum fluctuations as well as the Poynting vector due to the thermal fluctuations a stochastic quantity. It turns out that the standard deviation of the Casimir-Polder force excerted on a test particle is on the same order of magnitude as the mean value of the Casimir-Polder force itself~\cite{WuEtAl2002}. Since both phenomena --- TR and Casimir force --- are just two sides of the same medal, resulting from thermal and vacuum fluctuations of the electromagnetic field, such similarties could have been expected. However, as shown in Ref.~\cite{BiehsFluct} the standard deviation of the HF can become very large in the near-field regime which is a property not found for Casimir forces.

The aim of this letter is to make a first important step into the development of a general higher-order coherence theory for near-field TR within the framework of macroscopic quantum electrodynamics~\cite{MQED} which is a fully quantum version of Rytov's fluctuational electrodynamics~\cite{Rytov} mainly used in the field of near-field TR. To this end, we derive a general trace formula for the variance of the HF between two 
arbitrarily shaped objects immersed in an arbitrary environment. Therefore, we use the scattering approach which
describes the optical properties of the objects by their T-operators including non-reciprocal objects and environments
as needed in magneto-optical systems, for instance.  The derived expression complements the trace formulas for the
mean HF~\cite{KruegerEtAl2011} and allows for studying the fluctuations of the radiative HF around 
its mean value for arbitrarily shaped objects of any size for arbitrary distances, 
as long as the macroscopic approach is valid. Therefore, our general expression renders it possible to study heat 
flux fluctuations in far- and near-field regime for any two objects, which can have super- or sub-wavelength size. This work will pave the way for future theoretical studies of higher-order coherence theory for near-field TR and enhanced experimental setups for accessing these higher-order properties.

%%%%%%%%%%%%%%%%%%%%%%%%%%%%%%%%%%%%%%%%%%%%%%%%%%%%%%%%%%%%%%%%%%%%%%%%%%%%%%%%%%%%
%
% General expressions for heat flux between two objects 
%
%%%%%%%%%%%%%%%%%%%%%%%%%%%%%%%%%%%%%%%%%%%%%%%%%%%%%%%%%%%%%%%%%%%%%%%%%%%%%%%%%%%%

\section{General HF Expression}

Let us first start with the expression for the mean HF between two objects $\alpha$ and $\beta$ with a volume $V_\alpha$ and $V_\beta$ immersed in an arbitrary environment, assuming that the objects and environment are in local equilibrium at fixed temperatures $T_\alpha$, $T_\beta$, and $T_b$. In order to determine the HF between the objects $\alpha$ and $\beta$, one can determine the amount of heat dissipated in object $\alpha$, for instance, which is determined by the work done by the total field on the total currents inside the partice, i.e.\ by 
\begin{equation}
   \llangle H^\alpha \rrangle = \llangle[\Big] \sum_i \int_{V_\alpha}\!\!\!\rd^3 \mathbf{r} \,  \bigl\{ E_i (\mathbf{r},t) , J_{i,\alpha}(\mathbf{r},t) \bigr\}_{\rm S}\rrangle[\Big]
\label{Eq:General}
\end{equation}
where we use the symetrically ordered expression for the electric field and current operators as indicated by the anticommutator $\{E_i,J_i\}_S = (E_i J_i + J_i E_i)/2$. As detailed in Ref.~\cite{GreenKubo2} by using the scattering approach and the fluctuation dissipation theorem~\cite{Bimonte2009,KruegerEtAl2011,Messina2011,KruegerEtAl2012} for evaluating the averages, assuming that the environmental fields and the source currents in both object are in local thermal equilibrium, one arrives at a compact expression for the dissipated heat in object $\alpha$. It reads 
\begin{equation}
\begin{split}
   \llangle H^\alpha \rrangle &= 3 \int_0^\infty\!\!\frac{\rd \omega}{2 \pi}  \, \hbar \omega \bigl[ (n_\alpha(\omega) - n_b(\omega)) \mathcal{T}_{1}^\alpha (\omega) \\
          &\,\qquad\qquad\qquad  +  (n_\beta (\omega) - n_b(\omega)) \mathcal{T}_{2}^\alpha (\omega) \bigr] 
\end{split}
\label{Eq:Halpha}
\end{equation}
introducing the mean photonic occupation number $n_{\alpha/\beta/b}(\omega) = 1/(\exp(\hbar \omega/\kb T_{\alpha/\beta/b}) - 1)$ at temperature $T_{\alpha/\beta/b}$ with the Planck constant $\hbar$, Boltzmann constant $\kb$, and the transmission coefficients $\mathcal{T}_{1}^\alpha$ and $\mathcal{T}_{2}^\alpha$. Note that the overall sign is chosen such that the dissipated heat into object $\alpha$ is positive. From the above expression it is clear that, when taking $T_\alpha = T_b$, then there can only be a HF from object $\beta$ to object $\alpha$ so that the inter-object HF is 
\begin{equation}
   \llangle  H_{\beta \leftrightarrow \alpha} \rrangle = 3 \int_0^\infty\!\!\frac{\rd \omega}{2 \pi}  \, \hbar \omega (n_\beta - n_\alpha)  \mathcal{T}_{\beta \rightarrow \alpha}.
\label{Eq:MeanFlux}
\end{equation}
Another way to obtain this expression is to start from Eq.~(\ref{Eq:General}) and only to use the field and induced currents generated by the thermal sources in object $\beta$ at temperature $T_\beta$ giving the HF $\llangle  H_{\beta \rightarrow \alpha} \rrangle(T_\beta) \equiv \llangle \tilde{H}_\alpha \rrangle $ from object $\beta$ to $\alpha$ which coincides with Eq.~(\ref{Eq:MeanFlux}) for $n_\alpha = 0$. The backflow, due to the fact that object $\alpha$ has a temperature $T_\alpha$, can be calculated in the same manner just by assuming that the sources in $\beta$ have temperature $T_\alpha$ so that the amount of backflow is $\llangle  H_{\alpha \rightarrow \beta} \rrangle(T_\alpha) \equiv \llangle \tilde{H}_\beta \rrangle$ and the overall exchanged power results in $\llangle  H_{\beta \leftrightarrow \alpha} \rrangle = \llangle  \tilde{H}_{\alpha} \rrangle - \llangle  \tilde{H}_{\beta} \rrangle$. The transmission coefficient of interest $\mathcal{T}_{\beta \rightarrow \alpha} =  - \mathcal{T}_{2}^\alpha $ can be expressed as~\cite{GreenKubo2}
\begin{equation}
   \mathcal{T}_{\beta \rightarrow \alpha} = - \frac{4}{3}\Im {\rm Tr}\bigl[\mathds{P}\bigr]
\label{Eq:TC}
\end{equation}
within the scattering approach, where the trace is not the usual trace but the operator trace
\begin{equation}
   {\rm Tr}\bigl[ \mathds{P} \bigr] = \sum_i \int\!\!\rd^3 r\, \langle \mathbf{r} | \mathds{P}_{ii} | \mathbf{r} \rangle.
\end{equation}
By this definition the trace has the usual trace properties. The operator $\mathds{P}$ can be expressed as
\begin{equation}
  \mathds{P} = \mathds{O}_\beta  \mathds{G} \boldsymbol{\chi}_\beta  \mathds{G}^\dagger \mathds{D}_{\beta\alpha}^\dagger \mathds{T}_{\alpha}^\dagger,
\end{equation}
where $\mathds{G}$ is the operator of the Green's function taking the geometry of the environment into account, $\mathds{D}_{\alpha\beta} = \bigl( \mathds{1} - \mathds{G} \mathds{T}_\alpha \mathds{G} \mathds{T}_\beta \bigr)^{-1}$ is a Fabry-Perot like term taking the multiple interactions between object $\alpha$ and $\beta$ into account,  $\mathds{O}_\beta = \bigl(\mathds{1} + \mathds{G} \mathds{T}_\alpha\bigr) \mathds{D}_{\beta\alpha}$, and $\mathds{T}_{\alpha/\beta}$ is the T-operator of both scatterers $\alpha$ and $\beta$ determined by the Lippmann-Schwinger equation. Finally, $\boldsymbol{\chi}_\beta$ is the generalized susceptibility of object $\beta$ defined as~\cite{GreenKubo2} 
\begin{equation}
  \boldsymbol{\chi}_{\beta} = \frac{\mathds{T}_{\beta} - \mathds{T}_{\beta}^\dagger}{2 \ri} - \mathds{T}_{\beta} \frac{\mathds{G} - \mathds{G}^\dagger}{2 \ri} \mathds{T}_{\beta}^\dagger.
\end{equation}
In this way, the HF is expressed in a very general manner for objects and environment of any geometry which is fully contained in the T-operators and the Green function.

%%%%%%%%%%%%%%%%%%%%%%%%%%%%%%%%%%%%%%%%%%%%%%%%%%%%%%%%%%%%%%%%%%%%%%%%%%%%%%%%%%%%
%
%  General expression of the variance in the heat flux
%
%%%%%%%%%%%%%%%%%%%%%%%%%%%%%%%%%%%%%%%%%%%%%%%%%%%%%%%%%%%%%%%%%%%%%%%%%%%%%%%%%%%%

\section{General Expression for the Variance of the HF}

Now, the variance of the HF from object $\beta$ to object $\alpha$ is defined as 
\begin{equation}
\begin{split}
  {\rm Var} &:= \llangle (\tilde{H}_{\alpha} -\tilde{H_\beta})^2 \rrangle  - (\llangle \tilde{H}_{\alpha} \rrangle - \llangle \tilde{H}_{\beta} \rrangle )^2 \\
            &=  \llangle \tilde{H}_{\alpha}^2 \rrangle - \llangle \tilde{H}_{\alpha} \rrangle^2 +  \llangle \tilde{H}_{\beta}^2 \rrangle - \llangle \tilde{H}_{\beta} \rrangle^2 
\end{split}
\end{equation}
assuming that the sources at $T_\alpha$ and $T_\beta$ are statistically independent so that $\llangle \tilde{H}_{\alpha} \tilde{H}_{\beta} \rrangle = \llangle \tilde{H}_{\alpha} \rrangle \llangle \tilde{H}_{\beta} \rrangle$. To evaluate this expression we need to evaluate the first term only, which can be done by assuming that the field and current operators of TR fulfull the typical Gauss property of TR~\cite{Goodman}. Then after a lengthy and tedious calculation\cite{SM}  we arrive at the compact trace formula for the variance of the HF given as
\begin{equation}
  \begin{split}
     {\rm Var} &= \! \int_0^\infty\!\!\frac{\rd \omega}{2 \pi} \!\int_0^\infty\!\!\frac{\rd \omega'}{2 \pi} \bigl[ n_\alpha (\omega) n_\alpha(\omega') + n_\beta(\omega) n_\beta(\omega') \bigr] \\
               &\quad \times 16 \hbar^2 \omega \omega' \biggl( \frac{\omega}{\omega'} \Tr \bigl[ \Re\bigl(\mathds{M}(\omega) \bigr)  \Re\bigl(\mathds{N}(\omega') \bigr) \bigr]     \\         
               &\qquad  + \Tr \bigl[ \Im\bigl(\mathds{P}(\omega) \bigr)  \Im\bigl(\mathds{P}(\omega') \bigr) \bigr] \biggr)
  \end{split}
\label{Eq:HFF}
\end{equation}
with the operators 
\begin{align}
   \mathds{M} &= \mathds{O}_\beta  \mathds{G} \boldsymbol{\chi}_\beta  \mathds{G}^\dagger \mathds{O}^\dagger_\beta, \\
   \mathds{N} &= \mathds{T}_\alpha \mathds{D}_{\beta\alpha} \mathds{G} \boldsymbol{\chi}_\beta  \mathds{G}^\dagger \mathds{D}_{\beta\alpha}^\dagger \mathds{T}^\dagger_\alpha.
\end{align}
This general expression for the variance of the HF between two arbitrarily shaped objects $\alpha$ and $\beta$ immersed in an environment of arbitrary geometry is the main result of this work. It is valid for far- and near-field TR and any size of the objects, given that the distances and sizes are covered by macroscopic electrodynamics. Note that, because of our interest in the HF fluctuations, we have only taken thermal fluctuations into account, i.e.\ we have omitted any vacuum fluctuational part in the calculations. This is analogous to the treatment of the Casimir force fluctuations~\cite{WuEtAl2002} where the thermal part is completely neglected right from the start because the interest lies in the vacuum fluctuations, only. In general, the thermal and vacuum fluctuations contribute both to the HF and Casimir force fluctuations. For the interested reader we provide the corresponding expression for the vacuum contribution to the variance of the HF in the supplemental information~\cite{SM}. Furthermore, for reciprocal objects and environment one has $\mathds{T}^t_{\alpha/\beta} = \mathds{T}_{\alpha/\beta}$ and $\mathds{G}^t = \mathds{G}$ and therefore also $\mathds{M}^t = \mathds{M}$ and $\mathds{N}^t = \mathds{N}$ so that in this case the hermitian operators $\mathds{M}$ and $\mathds{N}$ fulfill $\Re(\mathds{M}) = \mathds{M}$ and $\Re(\mathds{N}) = \mathds{N}$. In this case the real part in the first trace expression can be omitted. % and the trace formula for the variance simplifies. 
%\begin{equation}
%  \begin{split}
%     {\rm Var} &= \! \int_0^\infty\!\!\frac{\rd \omega}{2 \pi} \!\int_0^\infty\!\!\frac{\rd \omega'}{2 \pi} \bigl[ n_\alpha (\omega) n_\alpha(\omega') + n_\beta(\omega) n_\beta(\omega') \bigr] \\
%               &\quad \times 16 \biggl( \hbar^2 \omega^2 \Tr \bigl[ \mathds{M}(\omega)\mathds{N}(\omega') \bigr] \\
%               &\qquad + \hbar^2 \omega \omega' \Tr \bigl[ \Im\bigl(\mathds{P}(\omega) \bigr)  \Im\bigl(\mathds{P}(\omega') \bigr) \bigr] \biggr).
%  \end{split}
%\end{equation}

From the general trace formula in Eq.~(\ref{Eq:HFF}) it is evident that even if we have a zero mean HF, i.e.\ if we have established global equilibrium by setting $T_\alpha = T_\beta = T_b$, then there is still a variance of the HF so that even in global equilibrium we have HF fluctuations.  Furthermore, the fluctuations are symmetric with respect to the temperatures of object $\alpha$ or $\beta$ in the sense that the fluctuations are proportional to $\bigl[ n_\alpha (\omega) n_\alpha(\omega') + n_\beta(\omega) n_\beta(\omega') \bigr]$ and therefore a temperature exchange $T_\alpha \rightarrow T_\beta$ leaves the fluctuations invariant whereas this exchange changes the sign of the mean HF. Of course, this symmetry is only valid if the temperature dependence of the material properties can be neglected. For phase-change materials like VO$_2$, for instance, the amplitude of HF changes dramatically by interchanging $T_\alpha \leftrightarrow T_\beta$ when working around the phase change temperature of 340K. This diode effect~\cite{PBASAB2013,LipingWang,ItoEtAl,FiorinoEtAl2018} will also have an impact on the amplitude of the fluctuations.

%%%%%%%%%%%%%%%%%%%%%%%%%%%%%%%%%%%%%%%%%%%%%%%%%%%%%%%%%%%%%%%%%%%%%%%%%%%%%%%%%%%%
%
% Heat Flux between two Nanoparticles
%
%%%%%%%%%%%%%%%%%%%%%%%%%%%%%%%%%%%%%%%%%%%%%%%%%%%%%%%%%%%%%%%%%%%%%%%%%%%%%%%%%%%%

\section{HF and its Variance between two NPs}

We now discuss the HF fluctuations for the special case of radiative heat transfer between two identical spherical nano-particles with radius $R$ and an interparticle distance $d$. We assume that the radius is much smaller than the thermal wavelength and that the distance $d$ is larger than $3R$. In this case the dipole model is valid and the T-operator in position space simplifies to~\cite{Asheichyk}
\begin{equation}
  \mathds{T}_{\alpha} (\mathbf{r},\mathbf{r}') = k_0^2 \alpha_{\rm np} \mathds{1} \delta(\mathbf{r} - \mathbf{r}_{\alpha})  \delta(\mathbf{r}' - \mathbf{r}_{\alpha}) 
	\label{Eq:Talpha}
\end{equation}
introducing the polarizability $\alpha_{\rm np}$ of the NP $\alpha$ at position $\mathbf{r}_{\alpha}$ and the vacuum wavenumber $k_0 = \omega/c$ with the light velocity in vacuum $c$. Of course, a similar expression is valid for $\mathds{T}_{\beta}$. As a consequence, the generalized polarizability of particle $\alpha$ has the form
\begin{equation}
  \boldsymbol{\chi}_\alpha = k_0^2 \bigl[ \Im(\alpha_{\rm np}) - k_0^2 |\alpha_{\rm np}|^2 \Im(G) \bigr] \mathds{1} \equiv \chi_\alpha \mathds{1}
\end{equation}
and is equal to the generalized polarizability of particle $\beta$, i.e.\ $\chi_\beta = \chi_\alpha$. Note that for equal arguments $\mathds{G}(\mathbf{r}_\beta, \mathbf{r}_\beta) = \mathds{G}(\mathbf{r}_\alpha, \mathbf{r}_\alpha) \equiv \mathds{1} G$, because of the isotropy and homogeneity of vacuum. Furthermore, all the operators $\mathds{G}(\mathbf{r},\mathbf{r'})$, $\mathds{D}(\mathbf{r},\mathbf{r}')$, etc.\ are diagonal so that it is very simple to determine the operator traces needed to evaluate the mean HF in (\ref{Eq:MeanFlux}) and variance of the HF in (\ref{Eq:HFF}). {Furthermore, we only consider electric polarizabilities which are relevant for dielectric materials in the infrared, whereas for metallic NPs also magnetic polarizabilites need to be taken into account~\cite{POCetal2008,AMetal2012}.} 

The transmission coefficent in Eq.~(\ref{Eq:TC}) can be expressed as the sum of two transversal and one longitudinal mode
\begin{equation}
   \mathcal{T}_{\beta \rightarrow \alpha} = \frac{1}{3} \bigl[ 2 \mathcal{T}_\perp (\omega) + \mathcal{T}_\parallel (\omega) \bigr].
\end{equation}
The transmission coefficients of the transversal and longitudinal mode are given by
\begin{equation}
  \mathcal{T}_{\perp/\parallel} = \frac{4 \chi_\beta \chi_\alpha |G_{\perp/\parallel}|^2}{|1 - k_0^4 \alpha_{\rm np}^2 G_{\perp/\parallel}^2|^2}. 
\end{equation}
The terms $G_{\perp/\parallel}(d)$ are just the components of the Green's function in vacuum perpendicular and parallel to the axis connecting particle $\alpha$ and $\beta$ which are at distance $d = |\mathbf{r}_\alpha - \mathbf{r}_\beta| > 0$, i.e.\ $\mathds{G}(\mathbf{r}_\alpha,\mathbf{r}_\beta) = G_{\perp}(d) \mathbf{e}_\perp~\otimes~\mathbf{e}_\perp +  G_{\parallel}(d) \mathbf{e}_\parallel\otimes \mathbf{e}_\parallel$ with unit vectors $ \mathbf{e}_{\perp/\parallel}$ perpendicular and parallel to the axis connecting the two NPs. They can be written as
\begin{align}
   G_{\perp} (d)    &=\frac{e^{\ri k_0 d}}{4\pi d} \biggl[ \frac{k_0^2 d^2 + \ri k_0d-1}{k_0^2 d^2} \biggr], \\
   G_{\parallel} (d) &=\frac{e^{\ri k_0 d}}{4\pi d} \biggl[ \frac{-2\ri k_0d+2}{k_0^2 d^2} \biggr].
\end{align}
Hence, the mean HF between the two NPs is
\begin{equation}
  \llangle H_{\beta \leftrightarrow \alpha} \rrangle = \int_0^\infty\!\!\frac{\rd \omega}{2 \pi}  \,  \hbar \omega (n_\beta - n_\alpha)  \bigl[ 2 \mathcal{T}_\perp (\omega) + \mathcal{T}_\parallel (\omega) \bigr]
\end{equation}
as, for example, already derived in a different form by Volokitin (see Ref.~\cite{Volokitin} and Refs.\ therein). Since the transmission coefficients $\mathcal{T}_{\perp/\parallel}$ are proportional to $|G_{\perp/\parallel}|^2$, the HF scales like $1/d^6$ in the near-field regime and in the far-field region like $1/d^2$ as is well known for dipole-dipole interactions. But note that due to the denominator of the transmission coefficients $\mathcal{T}_{\perp/\parallel}$ the HF is bounded and does not diverge as $d \rightarrow 0$. In the near-field region $k_0 d \ll 1$, it is easy to show that $\mathcal{T}_{\perp/\parallel} \leq 1$. Thus, the maximum HF is obtained when $\mathcal{T}_{\perp/\parallel} = 1$ for all frequencies resulting in $\llangle H_{\beta \leftrightarrow \alpha} \rrangle^{\rm max} = 3 \pi \kb^2/(12 \hbar) (T_\beta^2 - T_\alpha^2)$. This means that each of the three modes contributes to the conductance at most one quantum of thermal conductance as is known for the interparticle heat transfer even in many-particle systems~\cite{PBAEtAl2011,KathmannEtAl2018}.

Inserting the T-operator of the NPs into the trace terms in the expression for the variance of the HF in Eq.~(\ref{Eq:HFF}), we find
%\begin{equation}
%  \Tr \bigl[ \Im\bigl(\mathds{P}(\omega) \bigr)  \Im\bigl(\mathds{P}(\omega') \bigr) \bigr] = \frac{2 T_\perp (\omega) T_\perp (\omega') 
%                              +  T_\parallel (\omega) T_\parallel (\omega')}{16} 
%\end{equation}
%and because of the reciprocity of the particles and the environment we obtain 
%\begin{equation}
%\begin{split}
%  \Tr \bigl[\mathds{M}(\omega) \mathds{N}(\omega') \bigr] \! &= \! \frac{2 T_\perp (\omega) T_\perp (\omega') +  T_\parallel (\omega) T_\parallel (\omega')}{16 \chi_\alpha (\omega) \chi_\alpha (\omega')} \\
%                      &\quad \times |\alpha_{\rm np}(\omega')|^2 |1 + G k_0^2 \alpha_{\rm np}(\omega)|^2
%\end{split}
%\end{equation}
%so that the variance is 
\begin{equation}
  \begin{split}
    \!\! {\rm Var}\! &= \!\! \int_0^\infty\!\!\frac{\rd \omega}{2 \pi} \!\int_0^\infty\!\!\frac{\rd \omega'}{2 \pi} \bigl[ n_\alpha (\omega) n_\alpha(\omega') + n_\beta(\omega) n_\beta(\omega') \bigr] \\
               &\quad \times \!\! \hbar^2 \omega \omega' \bigl[ 2 T_\perp (\omega) T_\perp (\omega') +  T_\parallel (\omega) T_\parallel (\omega') \bigr] \\
               &\quad \times \!\! \biggl( \frac{\omega \omega'^3}{c^4} \frac{ |1 + G k_0^2 \alpha_{\rm np}(\omega)|^2 |\alpha_{\rm np}(\omega')|^2}{\chi_\alpha (\omega) \chi_\alpha (\omega')} + 1 \biggr).
  \end{split}
\label{Eq:HFFNP}
\end{equation}
From this expression it is already obvious that the variance scales in the near-field regime like $1/d^{12}$ and in far-field regime like $1/d^4$. Hence, the standard deviation of the HF has the same distance dependence as the mean HF in these extreme limits of distances much smaller or much larger than the thermal wavelength. As can be seen from this expression, we cannot derive a general maximal variance as we have done for the mean HF by setting $\mathcal{T}_{\perp/\parallel} = 1$ for all frequencies, because it depends on the polarizability and not on the transmission coefficients only. Nonetheless, we can evaluate the maximum contribution of the second term in the round brackets giving ${\rm Var}_{2,\rm max} := 3 \bigl( \pi \kb^2/ (12 \hbar) \bigr)^2 (T_\alpha^4 + T_\beta^4) $ which is similarly to the maxumum HF also universal. Hence, when setting $T_\alpha = 0\,{\rm K}$ then the maximal HF received by object $\alpha$ is $\llangle H_{\beta \leftrightarrow \alpha} \rrangle^{\rm max} = 3 \pi \kb^2 T_\beta^2 /(12 \hbar)$, and the maximal standard deviation by the second term of the variance is $\sigma_{2, \rm max} = \sqrt{{\rm Var}_{2,\rm max}} = \sqrt{3} \pi \kb^2 T_\beta^2 /(12 \hbar)$ so that $\sigma_{2,\rm max} = \llangle H_{\beta \leftrightarrow \alpha} \rrangle^{\rm max}/\sqrt{3}$ which is reminiscent of the blackbody result $\sigma_I = \langle I \rangle / \sqrt{2}$ where $\langle I \rangle$ is the intensity of a beam of unpolarized TR~\cite{MandelWolf}.

It can be further noted that the mean HF only contains the quantity $\Im(G)$, whereas the variance includes $\Re(G)$ and $\Im(G)$. It is well known that $\Im(G) = k_0/6\pi$ and that $\Re(G)$ is divergent which is an artefact of the point-like treatment of the NPs~\cite{Bladel} which is usually circumvented by introducing some kind of dressed polarizability or by simply neglecting the radiation correction terms which is equivalent to setting $G = 0$. In order to be able to numerically evaluate the expression of the variance without the hurdles of using the correct choice of dressed polarizabilities, as for example discussed in much detail for the discrete-dipole approximation in Ref.~\cite{Lakhtakia}, here we decide to use the prescription from~\cite{Yaghjian} where the finite size of the particle can be included by replacing $\mathds{G}(\mathbf{r}_\alpha,\mathbf{r}_\alpha)$ by the volume average
\begin{equation}
	\langle \mathds{G} \rangle_V = \frac{1}{V} \int_V\!\!\rd^3 r'\, \mathds{G} (\mathbf{r_\alpha,\mathbf{r}'})
\label{Eq:Yaghjian}
\end{equation}	
over the particle volume $V$ giving $\langle \mathds{G} \rangle_V = - \mathds{1}/(4 \pi k_0^2 R^3)$ in the limit $k_0 R \rightarrow 0$ for a spherical NP as shown in Ref.~\cite{Yaghjian} and, hence, $G = - 1/(4 \pi k_0^2 R^3)$. This procedure and the result are similar to that used in Ref.~\cite{Mahanty1973} for determining the Casimir-Polder force. Within the same limit $k_0 R \ll 1$ we can neglect the radiation correction, because it has been already found in other works that it is negligibly small for configurations where the dipole model is valid~\cite{Messina,Herz2018}. Then the generalized susceptibilities are simply $\chi_\alpha = \chi_\beta = k_0^2 \Im(\alpha_{\rm np})$ 
% and the variance simplifies to
%\begin{equation}
%  \begin{split}
%     {\rm Var} &= \!\! \int_0^\infty\!\!\frac{\rd \omega}{2 \pi} \!\int_0^\infty\!\!\frac{\rd \omega'}{2 \pi} \bigl[ n_\alpha (\omega) n_\alpha(\omega') + n_\beta(\omega) n_\beta(\omega') \bigr] \\
%               &\quad \times \hbar^2 \omega \omega' \bigl[ 2 T_\perp (\omega) T_\perp (\omega') +  T_\parallel (\omega) T_\parallel (\omega') \bigr] \\
%               &\quad \times \biggl( \frac{\omega}{\omega'} \frac{ |\tilde{\alpha}(\omega')|^2}{\Im(\tilde{\alpha}(\omega)) \Im(\tilde{\alpha}(\omega'))} + 1 \biggr)
%  \end{split}
%\label{Eq:HFFNPnoradcor}
%\end{equation}
and the transmission coefficients simplify to
\begin{equation}
  \mathcal{T}_{\perp/\parallel} = \frac{4 k_0^4 \Im(\alpha_{\rm np})^2 |G_{\perp/\parallel}|^2}{|1 - k_0^4 \alpha_{\rm np}^2 G_{\perp/\parallel}^2|^2}. 
\end{equation}
Note, that in the near-field regime $k_0 d \ll 1$ the quantities $G_\perp \approx -1/(4 \pi k_0^2 d^3)$ and $G_\parallel \approx 2/(4 \pi k_0^2 d^3)$ are purely real so that the maximum value of $\mathcal{T}_{\perp/\parallel}$ is indeed equal to one and it is obtained if the condition $k_0^4 |\alpha_{\rm np}|^2 G_{\perp/\parallel}^2 = 1$ is fulfilled.  

\begin{figure}
  \includegraphics[width=0.45\textwidth]{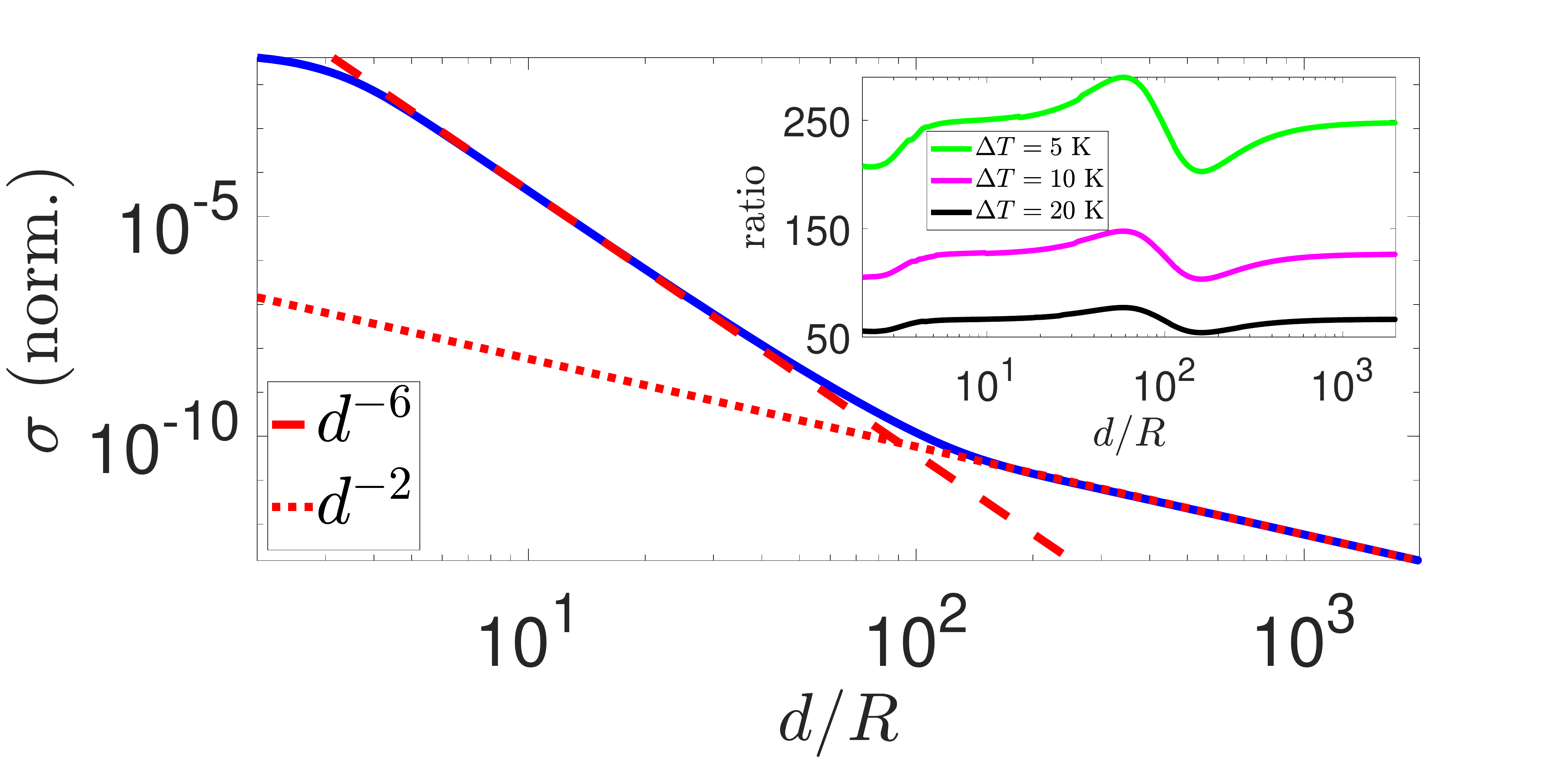}
	\caption{Standard deviation $\sigma = \sqrt{\rm Var}$ normalized to $\sigma_{2,{\rm max}} = \sqrt{3} \pi \kb^2/ (12 \hbar) \sqrt{T_\alpha^4 + T_\beta^4}$ for two SiC NPs with $25\,{\rm nm}$ radii held at temperatures $T_\alpha = 300\,{\rm K}$ and $T_\beta =  T_\alpha + \Delta T$ with $\Delta T = 20\,{\rm K}$. The $1/d^6$ and $1/d^2$ dependence of $\sigma$ in the near- and far-field regime can be clearly seen together with a saturation for $d = 2 R$. The inset shows the ratio of the standard deviation $\sigma$ and mean HF $\sigma / \llangle H_{\beta \leftrightarrow \alpha} \rrangle$ for different temperature differences $\Delta T$. }
  \label{Fig:Varianz}
\end{figure}

These expressions are now evaluated numerically using the polarizability $\alpha_{\rm np} = 4 \pi R^3 (\epsilon - 1)/(\epsilon + 2)$ with the particles permittivity $\epsilon$ for SiC~\cite{Palik}. In Fig.~\ref{Fig:Varianz} we show a plot of the standard deviation $\sigma = \sqrt{\rm Var}$ and the ratio  $\sigma/\llangle H_{\beta \leftrightarrow \alpha} \rrangle$ for two SiC NPs at $T_\alpha = 300\,{\rm K}$ and $T_\beta =  T_\alpha + \Delta T$. It can be seen that the standard deviation can be approximately 77 times larger than the mean value for $\Delta T = 20\,{\rm K}$ and about 290 times larger for  $\Delta T = 5\,{\rm K}$. We find numerically that for small $\Delta T \ll T_\alpha$ the ratio of the standard deviation and mean HF $\sigma / \llangle H_{\beta \leftrightarrow \alpha} \rrangle$ scales like $1/ \Delta T$. This is the same temperature dependence as for the ratio of $\sigma_{2,{\rm max}} = \sqrt{3} \pi \kb^2/ (12 \hbar) \sqrt{T_\alpha^4 + T_\beta^4}$ and the maximum HF $\llangle H_{\beta \leftrightarrow \alpha} \rrangle^{\rm max}$. Actually, the $1/ \Delta T$ scaling is a general feature because the variance or standard deviation goes to some constant value for $\Delta T \rightarrow 0$, whereas the HF goes linearly in $\Delta T$ to zero. As a consequence, the relative amplitude of the HF fluctuations becomes extremely large for small temperature differences, as pointed out recently for two semi-infinite materials~\cite{BiehsFluct}. We also find that the variance depends on the choice of $G$ in Eq.~(\ref{Eq:Yaghjian}). When neglecting the radiation correction, as it is usually done, by setting $G = 0$ the standard deviation becomes by a factor 6.26 smaller. More detailed studies on the impact of geometry, material properties, background, and zero fluctuations are needed to get more insight into the general nature of heat-flux fluctuations in near-field heat transfer and our trace formula provides the basis for such studies. 

%%%%%%%%%%%%%%%%%%%%%%%%%%%%%%%%%%%%%%%%%%%%%%%%%%%%%%%%%%%%%%%%%%%%%%%%%%%%%%%%%%%%
%
% particle interface 
%
%%%%%%%%%%%%%%%%%%%%%%%%%%%%%%%%%%%%%%%%%%%%%%%%%%%%%%%%%%%%%%%%%%%%%%%%%%%%%%%%%%%%
\section{HF and its Variance between a NP and a substrate and emission into free space}

{Finally, we discuss the HF and the variance of the HF for a spherical NP above a planar, isotropic and reciprocal substrate and the HF and its variance emitted by a NP into vacuum. These two situations might be relevant for tip-based or far-field experiments~\cite{Babuty2013,Kittel2008,Worbes2013,Jones2012,Weng2018,Komiyama2019}. To this end, we evaluate our general expressions by using the T-operator in Eq.~(\ref{Eq:Talpha}) for particle $\alpha$ at position $\mathbf{r}_\alpha = (0,0,d)^t$, i.e.\ the NP is in a distance $d$ above the substrate which occupies the halfspace $z < 0$. Furthermore, for convenience  we first consider as for example in Ref.~\cite{KruegerEtAl2012} the single-scattering approximation where $\mathds{D}_{\beta\alpha} \approx \mathds{1}$, since the contribution of the denominator can be neglected if $d > 3R$ as can be seen in Fig.~\ref{Fig:Varianz}. As detailed in~\cite{SM}, we then obtain
\begin{equation}
\begin{split}
	\llangle H_{\beta \leftrightarrow \alpha} \rrangle  &= \frac{4 \hbar}{c^2} \int_0^\infty \frac{\domega}{2 \pi} \omega^3 \left[ n_{\beta} (\omega) - n_{\alpha} (\omega) \right]\\
	                                                &\qquad\times\left( 2 \mathcal{C}_\perp (\omega) + \mathcal{C}_\parallel (\omega) \right) \Im\left(\alpha_\text{np} \right) 
\end{split}
\label{Eq:HFParticleSubstrate}
\end{equation}
for the mean power and 
\begin{equation}
\begin{split}
 \!\!\!\! \text{Var} & = \!\!\int_0^\infty \!\!\frac{\domega}{2 \pi} \!\! \int_0^\infty\!\! \frac{\domega'}{2 \pi}\left[ n_{\beta} (\omega) n_{\beta} (\omega') + n_{\alpha} (\omega) n_{\alpha} (\omega') \right] \\
& \quad \!\!\times\!\!  \frac{16 \hbar^2 \omega^3  \omega'^3}{c^4}  \left( 2 \mathcal{C}_\perp (\omega) \mathcal{C}_\perp (\omega') + \mathcal{C}_\parallel (\omega) \mathcal{C}_\parallel (\omega') \right) \\
	& \quad \!\!\times\!\! \Bigl[ \frac{\omega'}{\omega}\big|1 + k_0^2 \alpha_\text{np} G \big|^2  |\alpha_\text{np}'|^2 + \Im\left(\alpha_\text{np} \right) \Im\left(\alpha_\text{np}' \right) \Bigr] 
\end{split}
\label{Eq:VarianceParticleSubstrate}
\end{equation}
for the variance of the HF. The quantities $C_\perp(\omega)$ and $C_\parallel(\omega)$ are defined in Ref.~\cite{SM} and are connected to the electrical out-of-equilibrium local density of states $D^E(\omega;d)$~\cite{Doro} at the distance $d$ of the NP above the substrate. To be more specific, we find that $D^E (\omega;d) = \left( 2 \mathcal{C}_\perp (\omega) + \mathcal{C}_\parallel (\omega) \right) 2 \omega/\pi c^2$ so that our expression of the transferred power in Eq.~(\ref{Eq:HFParticleSubstrate}) reproduces the well-known HF expressions(see for instance, Ref.~\cite{Volokitin}). On the other hand, the expression for the variance is new and it is loosely speaking proportional to the square of the local density of states. Therefore we can already expect that again $\sigma/\llangle H_{\beta \leftrightarrow \alpha} \rrangle \approx {\rm const}$ in the near-field, but also far-field regime. Furthermore, in the near-field regime the HF scales like $1/d^3$ and the variance like $1/d^6$. }

\begin{figure}
  \includegraphics[width=0.45\textwidth]{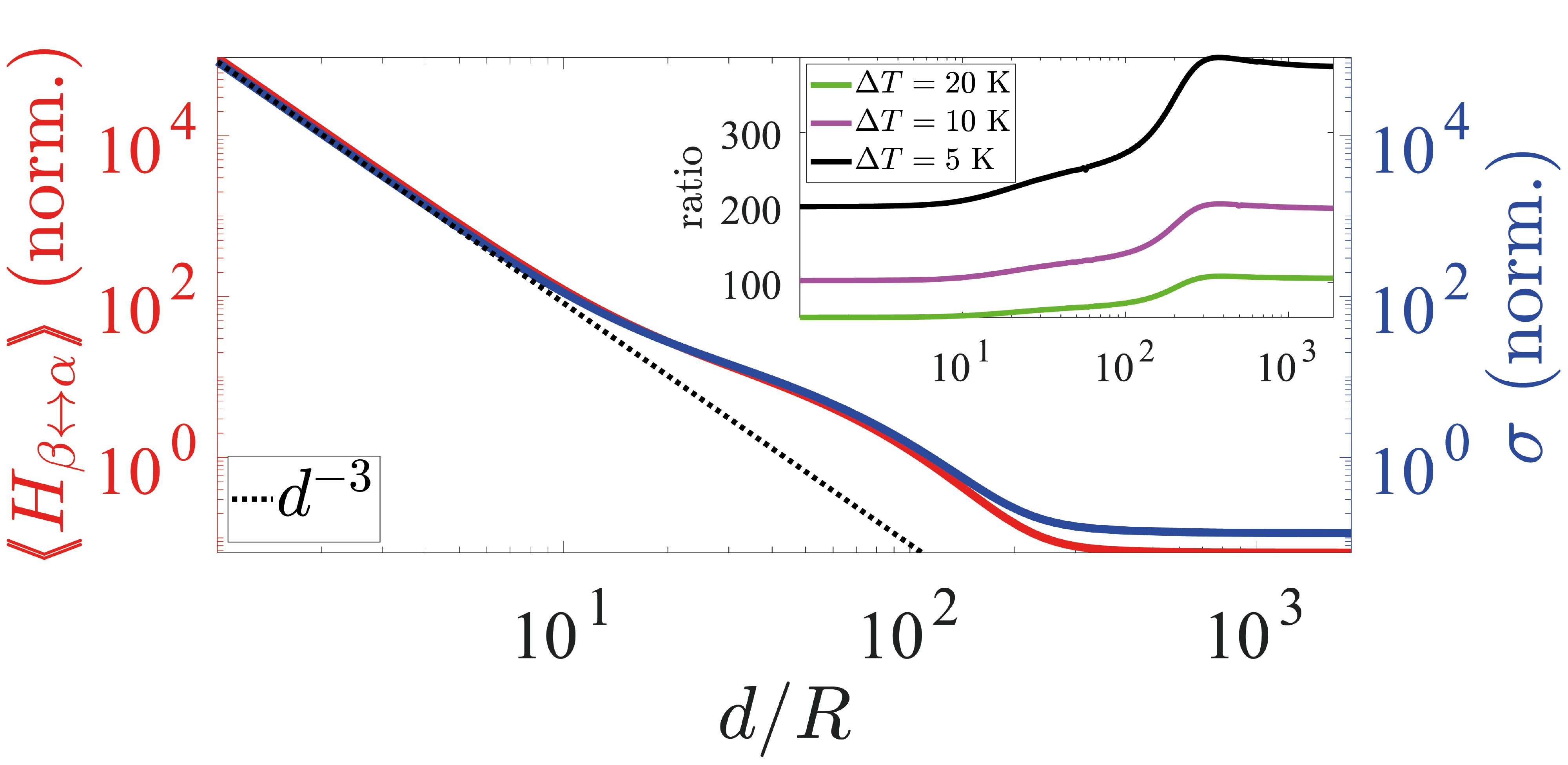}
	\caption{ Standard deviation $\sigma = \sqrt{\rm Var}$ and transferred power $\llangle H_{\beta \leftrightarrow \alpha} \rrangle$ for a SiC NPs with $25\,{\rm nm}$ radius held at temperature $T_\alpha = 300\,{\rm K}$ above a planar SiC substrate at $T_\beta =  T_\alpha + \Delta T$ with $\Delta T = 20\,{\rm K}$. The values are normalized to the corresponding quantities where the substrate is replaced by vacuum.   The $1/d^3$ dependence of both quantities stemming from the local density of states in the near-field regime can be clearly seen. The inset shows $\sigma / \llangle H_{\beta \leftrightarrow \alpha} \rrangle$ for different temperature differences $\Delta T$. }
  \label{Fig:VarianzSubstrate}
\end{figure}

From our expressions (\ref{Eq:HFParticleSubstrate}) and (\ref{Eq:VarianceParticleSubstrate}) we can easily obtain the HF and its variance received by a NP which is irradiated by a blackbody at temperature $T_\beta$ by replacing the substrate by vacuum. This corresponds also to the HF and its variance which is emitted by a NP into its surrounding and might therefore be relevant for far-field measurements of heat radiation and its variance of nano- or micron-sized particles. We find that when assuming $T_\alpha > T_\beta$, where $T_\beta$ is the temperature of the vacuum surrounding the particle, the HF of the particle into negative z-direction is
\begin{equation}
 \! \llangle H_{\alpha \leftrightarrow \beta} \rrangle  = \frac{\hbar}{\pi c^3} \int_0^\infty \!\!\!\frac{\domega}{2 \pi} \omega^4 \left[ n_{\alpha} (\omega) - n_{\beta} (\omega) \right] \Im\left(\alpha_\text{np} \right)
\end{equation}
with the variance 
\begin{equation}
\begin{split}
  \text{Var} & = \frac{\hbar^2}{3 \pi^2 c^6} \int_0^\infty \frac{\domega}{2 \pi}  \!\! \int_0^\infty \frac{\domega'}{2 \pi} \Im\left(\alpha_\text{np} \right) \Im\left(\alpha_\text{np}'\right)\\
	     &\quad\!\!\times \! \omega^4 \omega'^4 \left[ n_{\beta} (\omega) n_{\beta} (\omega') + n_{\alpha} (\omega) n_{\alpha} (\omega') \right]  \\
	& \quad \!\!\times \!\!\left[ \frac{\omega'}{\omega} \frac{\big|1 + k_0^2 \alpha_\text{np} G \big|^2 |\alpha_\text{np}'|^2}{\Im\left(\alpha_\text{np} \right) \Im\left(\alpha_\text{np}'\right) } + 1  \right] .
\end{split}
\end{equation}
Note again, that our HF formula reproduces well-known results~\cite{Herz2018}, whereas the variance has not been determined before. In addition, by comparing the second term of the variance $\text{Var}_2$ with the mean power, we have $\sigma_2 = \sqrt{\text{Var}_2} = \sqrt{3} \llangle H_{\alpha \leftrightarrow \beta} \rrangle$ for $T_\beta = 0\,{\rm K}$ similar to what we obtained for two NPs.

In Fig.~\ref{Fig:VarianzSubstrate} we show the mean power and the standard deviation $\sigma = \sqrt{\text{Var}}$ for a SiC NP at a distance $d$ above a SiC substrate normalized to the corresponding quantities for the emission of a SiC NP into free space. It can be seen that in the near-field regime both quantities scale like $1/d^3$ which is due to the fact that both quantities are connected to the local density of states. Furthermore, it is obvious that the ratio $\sigma = \sqrt{\text{Var}}/\llangle H_{\alpha \leftrightarrow \beta} \rrangle$ can be very large as also found for the HF between two NPs. It is interesting to note that in contrast to the case of two NPs the ratio $\sigma = \sqrt{\text{Var}}/\llangle H_{\alpha \leftrightarrow \beta} \rrangle$ is larger in the far-field regime than in the near-field regime.

%%%%%%%%%%%%%%%%%%%%%%%%%%%%%%%%%%%%%%%%%%%%%%%%%%%%%%%%%%%%%%%%%%%%%%%%%%%%%%%%%%%%
%
%  Conclusions
%
%%%%%%%%%%%%%%%%%%%%%%%%%%%%%%%%%%%%%%%%%%%%%%%%%%%%%%%%%%%%%%%%%%%%%%%%%%%%%%%%%%%%
\section{Conclusion}

In conclusion, we have derived a general trace formula for the variance of the HF between two arbitrarily shaped
objects and discussed its consequences for two spherical NPs. By this, we have made the first important step
towards a general higher-order coherence theory for near-field TR. We are convinced that this theory will 
not only bring new interesting insights into the fundamental properties of TR at the nanoscale, but it will
pave the way for experimental and theoretical tests of the Gaussian property and fluctuation theorems for near-field termal radiation. In principle, the HF fluctuations could be assessed by measuring a time series of the HF in steady state and evaluating the statistics. One should observe large deviations from the mean value with the standard deviation determined in this work. However, since the fluctuations of near-field TR and therefore also for the HF happen to be on a time scale which is determined by the coherence time which is in the near-field between femto- and pico-seconds~\cite{BiehsFluct}, the fluctuations are typically averaged out in conventional HF measurements. Hence, ultra-fast near-field measurement methods need to be developed in the future in order to verify the here made predictions. {On the theory side, further connection to experiments as done for the Casimir force~\cite{Barton1991,Jaekel1992,PAM1993} can be made by studying the temporal correlations and spectra of the Poynting vector starting from Eq.~(19) in~\cite{SM}.}

%%%%%%%%%%%%%%%%%%%%%%%%%%%%%%%%%%%%%%%%%%%%%%%%%%%%%%%%%%%%%%%%%%%%%%%%%%%%%%%%%%%%
%
% Acknowledgements
%
%%%%%%%%%%%%%%%%%%%%%%%%%%%%%%%%%%%%%%%%%%%%%%%%%%%%%%%%%%%%%%%%%%%%%%%%%%%%%%%%%%%%

S.-A.\ B. acknowledges support from Heisenberg Programme of the Deutsche Forschungsgemeinschaft (DFG, German Research Foundation) under the project No. 404073166.

%%%%%%%%%%%%%%%%%%%%%%%%%%%%%%%%%%%%%%%%%%%%%%%%%%%%%%%%%%%%%%%%%%%%%%%%%%%%%%%%%%%%
%
% Bibliography
%
%%%%%%%%%%%%%%%%%%%%%%%%%%%%%%%%%%%%%%%%%%%%%%%%%%%%%%%%%%%%%%%%%%%%%%%%%%%%%%%%%%%%

\end{document}